\documentclass[review,times]{elsarticle}
\makeatletter
\def\ps@pprintTitle{%
 \let\@oddhead\@empty
 \let\@evenhead\@empty
 \def\@oddfoot{\centerline{\thepage}}%
 \let\@evenfoot\@oddfoot}
\makeatother

\usepackage{lineno,hyperref}

\usepackage[labelsep=period,labelfont=bf]{caption} 
\usepackage{amsmath}

\usepackage{hyperref}
\usepackage[hyperref]{xcolor}
\hypersetup{
  colorlinks=true,
}

\begin{document}

\begin{frontmatter}

\title{On sound propagation in three-dimensional poroelastic field: correct field expressions}


\author[mymainaddress]{Hou Qiao}

\author[mymainaddress,mysecondaryaddress]{Zeng He}

\author[mymainaddress,mysecondaryaddress]{Wen Jiang\corref{mycorrespondingauthor}}
\cortext[mycorrespondingauthor]{Corresponding author}
\ead{wjiang@hust.edu.cn}

\author[mythirdaddress]{Weicai Peng}

\address[mymainaddress]{Department of Mechanics, Huazhong University of Science $\&$Technology, Wuhan, China}
\address[mysecondaryaddress]{Hubei Key Laboratory for Engineering Structural Analysis and Safety Assessment, Huazhong University of Science $\&$ Technology, Wuhan, China}
\address[mythirdaddress]{National Key Laboratory on Ship Vibration and Noise, China Ship Development and Design Center, Wuhan, China}

\begin{abstract}

The correct three-dimensional poroelastic field expressions, under monochromatic harmonic incidence, were revealed in this paper. They were confirmed satisfactory using the results got by other researchers. Former incomplete expressions were found to overestimate the random sound transmission loss and lead to division-by-zero, though their results were amply acceptable.

\end{abstract}


\end{frontmatter}


\section{Introduction}

Recently, much interest has been shown on modeling sound absorbing materials, especially poroelastic materials, i.e., porous materials which have an elastic frame. The Biot theory \cite{Biot1956TJotASoA}, which is the most widely used modeling technique for these materials, is studied extensively during the last decades.

Despite the fruitful numerical developments \cite{Atalla1998TJotASoA,Atalla2006TJotASoA,Schanz2018EAwBE} in Biot theory, the related theoretical study is scarce and mainly about the two-dimensional (2D) problems \cite{Deresiewicz1962BotSSoA,Bolton1996JoSaV}. Closed-form 2D poroelastic field expressions, which are among the most notable progress, were obtained by Bolton \cite{Bolton1996JoSaV} using the simplifications of Deresiewicz \cite{Deresiewicz1962BotSSoA} and Allard \cite{Allard1989JoAP}. When it comes to the three-dimensional (3D) case, the theoretical development is insufficient until the work by Zhou \cite{Zhou2013JoSaV}, where he obtained the 3D poroelastic field expressions. These field expressions have been cited extensively \cite{Liu2015JoSaV,Talebitooti2016JoSaV} since then. 

However, the 3D poroelastic field expressions are not sufficiently revealed by now. For example, in the work by Zhou \cite{Zhou2013JoSaV,Zhou2013AA}, he used an inappropriate hypothesis (refer to Equation (13) in \cite{Zhou2013AA}), which leads to an incomplete set of expressions. It can be erroneous if the correct expressions are not revealed. The aim of this discussion is to provide their formal derivation, and the correct expressions. The field expressions obtained here, with satisfying consistency with the theoretical results obtained by other researchers, are confirmed adequate in 3D poroelastic applications. 

\section{Derivations}
With the time dependence ${\rm e}^{{\rm j}\omega t}$ (${\rm j}=\sqrt{-1}$), the poroelastic equations formulated by solid displacement ${\bf u}^s$ and fluid displacement ${\bf u}^f$ are \cite{Allard2009}
\begin{eqnarray}
    -\omega^2\left({\rho}_{11}^{*}{\bf u}^{s}+{\rho}_{12}^{*} {\bf u}^{f}\right)=(A+N)\nabla\nabla\cdot{\bf u}^{s}+N\nabla^2{\bf u}^{s}+Q\nabla\nabla\cdot{\bf u}^{f}\notag\\
    -\omega^2\left({\rho}_{22}^{*}{\bf u}^{f}+{\rho}_{12}^{*}{\bf u}^{s}\right)=R\nabla\nabla\cdot{\bf u}^{f} +Q\nabla\nabla\cdot{\bf u}^{s}\label{eq:biot1-2}
\end{eqnarray}
where the porous parameters ${\rho}_{11}^{*},{\rho}_{12}^{*},{\rho}_{22}^{*},A,N,Q,R$ can be found in \cite{Bolton1996JoSaV,Zhou2013JoSaV,Qiao2018a}. When two scalar potentials $\varphi_s=\nabla\cdot{\bf u}^{s}, \varphi_f=\nabla\cdot{\bf u}^{f}$ and two vector potentials ${\bf \Psi}_s=\nabla\times{\bf u}^{s}, {\bf \Psi}_f=\nabla\times{\bf u}^{f}$ are introduced, the above equations can be reduced to a fourth-order scalar wave equation and a second-order vector equation \cite{Bolton1996JoSaV}. The three propagation wavenumbers corresponding to the wave equations are
\begin{eqnarray}
\{k_{1}^2,k_{2}^2\}=A_1/2\pm\sqrt{A_1^2/4-A_2},\ k_3^2=\omega^2/N({\rho}_{11}^{*}-{\rho}_{12}^{*}{\rho}_{12}^{*}/{\rho}_{22}^{*}) \label{eq:biot-k123}
\end{eqnarray}
where $A_1=\omega^2({\rho}_{11}^{*}R-2{\rho}_{12}^{*}Q+{\rho}_{22}^{*}P)/(PR-Q^2)$, $A_2=\omega^4({\rho}_{11}^{*}{\rho}_{22}^{*}-{\rho}_{12}^{*}{\rho}_{12}^{*})/(PR-Q^2)$, $P=A+2N$; $k_{1},k_{2}$ correspond to the scalar wave equation, while $k_{3}$ correspond to the vector wave equation.

When a 3D monochromatic incident wave $\varPhi_i= {\rm e}^{{\rm j}\omega t-{\rm j} \bf{k} \bf{r}}$, where ${\bf k}=(k_x,k_y,k_z)$, ${\bf r}=(x,y,z)$, transmits through the porous domain along its normal direction $z$ (the time dependence ${\rm e}^{{\rm j}\omega t}$ is omitted henceforth), the solid phase strain $e_s=\varphi_s$ and ${\bf \Omega}_s={\bf \Psi}_s$ are
\begin{eqnarray}
	e_s={\rm e}^{-{\rm j}(k_x x+k_y y)}\left(C_1 {\rm e}^{-{\rm j} k_{1z} z}+C_2 {\rm e}^{{\rm j} k_{1z} z}+C_3 {\rm e}^{-{\rm j} k_{2z} z}+C_4 {\rm e}^{{\rm j} k_{2z} z} \right)\label{eq:biot_es}\\
	\begin{aligned}
	{\bf \Omega}_s={\rm e}^{-{\rm j}(k_x x+k_y y)}& \bigg[\left(C_5 {\rm e}^{-{\rm j} k_{3z} z} +C_6 {\rm e}^{{\rm j} k_{3z} z} \right){\bf i}_x + \left(C_7 {\rm e}^{-{\rm j} k_{3z} z}+C_8 {\rm e}^{{\rm j} k_{3z} z} \right){\bf i}_y\\&+ \left(C_9 {\rm e}^{-{\rm j} k_{3z} z}+C_{10} {\rm e}^{{\rm j} k_{3z} z} \right){\bf i}_z \bigg]
	\end{aligned}\label{eq:biot_ws}
\end{eqnarray}
where $k_{iz}=\sqrt{k_i^2-k_x^2-k_y^2}$ ($i$=1,2,3); ${\bf i}_x,{\bf i}_y,{\bf i}_z$ are the unit vectors along x, y, and z; the unknown constants $C_1-C_{10}$ are not independent and they can be determined by proper boundary conditions. Subsequently, the fluid phase strain $e_f=\varphi_f$ and ${\bf \Omega}_f={\bf \Psi}_f$ are
\begin{eqnarray}
&e_f={\rm e}^{-{\rm j}(k_x x+k_y y)}\left(b_1 C_1 {\rm e}^{-{\rm j} k_{1z} z}+b_1 C_2 {\rm e}^{{\rm j} k_{1z} z}+b_2 C_3 {\rm e}^{-{\rm j} k_{2z} z}+b_2 C_4 {\rm e}^{{\rm j} k_{2z} z} \right)\label{eq:biot_ef}
\\
&{\bf \Omega}_f=g {\bf \Omega}_s\label{eq:biot_wf}
\end{eqnarray}
where $b_1,b_2,g$ can be found in \cite{Bolton1996JoSaV,Zhou2013JoSaV}.

Bolton \cite{Bolton1996JoSaV} pointed out that, in 2D case, all displacement field variables should contain contributions from all the six wave components. Accordingly, in the 3D case here, the components of the solid phase displacement ${\bf u}^s=(u_x^s,u_y^s,u_z^s)$ are
\begin{eqnarray}
\begin{aligned}
	u_x^s={\rm e}^{-{\rm j}(k_x x+k_y y)}&\bigg(D_1 {\rm e}^{-{\rm j} k_{1z} z}+D_2 {\rm e}^{{\rm j} k_{1z} z}+D_3 {\rm e}^{-{\rm j} k_{2z} z}\\ &+D_4 {\rm e}^{{\rm j} k_{2z} z}+D_5 {\rm e}^{-{\rm j} k_{3z} z}+D_6 {\rm e}^{{\rm j} k_{3z} z} \bigg)
\end{aligned}\label{eq:[biot_usx_3d]}\\
\begin{aligned}
	u_y^s={\rm e}^{-{\rm j}(k_x x+k_y y)}&\bigg(D_7 {\rm e}^{-{\rm j} k_{1z} z}+D_8 {\rm e}^{{\rm j} k_{1z} z}+D_9 {\rm e}^{-{\rm j} k_{2z} z}\\ &+D_{10} {\rm e}^{{\rm j} k_{2z} z}+D_{11} {\rm e}^{-{\rm j} k_{3z} z}+D_{12} {\rm e}^{{\rm j} k_{3z} z} \bigg)
\end{aligned}\label{eq:[biot_usy_3d]}\\
\begin{aligned}
	u_z^s={\rm e}^{-{\rm j}(k_x x+k_y y)}&\bigg(D_{13} {\rm e}^{-{\rm j} k_{1z} z}+D_{14} {\rm e}^{{\rm j} k_{1z} z}+D_{15} {\rm e}^{-{\rm j} k_{2z} z}\\ &+D_{16} {\rm e}^{{\rm j} k_{2z} z}+D_{17} {\rm e}^{-{\rm j} k_{3z} z}+D_{18} {\rm e}^{{\rm j} k_{3z} z} \bigg)
\end{aligned}\label{eq:[biot_usz_3d]}
\end{eqnarray}
where $D_1-D_{18}$ are the contribution coefficients of the six wave components to the displacement components. These coefficients are not independent and they are expressed by $C_1-C_{10}$ subsequently. Corresponding results of the fluid phase, which can be obtained in a similar manner as the solid phase, are not discussed in detail herein.

As the unknown constants $C_1-C_{10}$ are not independent, the condition $\nabla\cdot{\bf\Omega}^s=0$ \cite{Graff1975} which can uniquely determine ${\bf u}^s$ from $e_s$ and ${\bf \Omega}^s$, is used first to eliminate the dependent constants among them. Subsequently, three different cases are identified according to their incident wavenumbers. The first case, when the incident wave is not in the $x$-$z$ plane ($k_y\ne 0$), leads to
\begin{eqnarray}
C_7=-\frac{k_x}{k_y}C_5,\ C_8=-\frac{k_x}{k_y}C_6,\ C_9=C_{10}=0 \label{eq:coef_ky_ne_0}
\end{eqnarray}
The second case, when the incident wave is not in the $y$-$z$ plane ($k_x\ne 0$), leads to
\begin{eqnarray}	
C_5=-\frac{k_y}{k_x}C_7,\ C_6=-\frac{k_y}{k_x}C_8,\ C_9=C_{10}=0 \label{eq:coef_kx_ne_0}
\end{eqnarray}
The third case, when the incident wave is along the $z$ axis (normal incidence, $k_x=k_y=0$), leads to
\begin{eqnarray}
C_5=C_6=C_7=C_8=C_9=C_{10}=0 \label{eq:coef_kx_ky_0}
\end{eqnarray}
The above cases, to be brief, are denoted as incident case 1 ($k_y\ne 0$), incident case 2 ($k_x\ne 0$) and incident case 3 ($k_x=k_y=0$) in the following. These cases are not mutually exclusive; however, they are separated for mathematical convenience. It can be concluded from Eqs.(\ref{eq:coef_ky_ne_0})-(\ref{eq:coef_kx_ky_0}) that there are six independent constants at most in the 3D poroelastic field here.

According to Eqs.(\ref{eq:biot_es})-(\ref{eq:biot_ws}), (\ref{eq:[biot_usx_3d]})-(\ref{eq:coef_kx_ky_0}), the poroelastic field expressions can be expressed in closed form using four or six independent constants. Once the field expressions are obtained, the independent unknown constants, along with the poroelastic problem, can be solved by proper boundary conditions.

\subsection{Incident case 1}
When $k_y\ne 0$, according to Eq.(\ref{eq:coef_ky_ne_0}), the solid phase strain ${\bf \Omega}_s$ is
\begin{eqnarray}
{\bf \Omega}_s={\rm e}^{-{\rm j}(k_x x+k_y y)} \left(C_5 {\rm e}^{-{\rm j} k_{3z} z} +C_6 {\rm e}^{{\rm j} k_{3z} z} \right)\left({\bf i}_x-\frac{k_x}{k_y}{\bf i}_y\right)
\end{eqnarray}
Subsequently, the displacement ${\bf u} =[u_x^s,u_y^s,u_z^s,u_x^f,u_y^f,u_z^f]^T$ is
\begin{eqnarray}
{\bf u}={\rm e}^{-{\rm j}(k_x x+k_y y)} {\bf Y}_1 {\bf E} {\bf C}
\end{eqnarray}
where
\begin{eqnarray}
&{\bf E} =diag({\rm e}^{-{\rm j} k_{1z}^{m} z},{\rm e}^{{\rm j} k_{1z}^{m} z},{\rm e}^{-{\rm j} k_{2z}^{m} z},{\rm e}^{{\rm j} k_{2z}^{m} z},{\rm e}^{-{\rm j} k_{3z}^{m} z},{\rm e}^{{\rm j} k_{3z}^{m} z})\label{eq:matrix_e}\\
&{\bf C} =[C_1,C_2,C_3,C_4,C_5,C_6]^T\label{eq:vec_c}
\end{eqnarray}
Here, the matrix ${\bf E}$ is a $6\times6$ diagonal matrix. The non-zero elements of the coefficient matrix ${\bf Y}_1$ are provided in \ref{app:case1_exp}. 

\subsection{Incident case 2}
When $k_x\ne 0$, according to Eq.(\ref{eq:coef_kx_ne_0}), the solid phase strain ${\bf \Omega}_s$ is
\begin{eqnarray}
{\bf \Omega}_s={\rm e}^{-{\rm j}(k_x x+k_y y)} \left(C_5 {\rm e}^{-{\rm j} k_{3z} z} +C_6 {\rm e}^{{\rm j} k_{3z} z} \right)\left({\bf i}_y-\frac{k_y}{k_x}{\bf i}_x\right)
\end{eqnarray}
Subsequently, the displacement ${\bf u}=[u_x^s,u_y^s,u_z^s,u_x^f,u_y^f,u_z^f]^T$ is
\begin{eqnarray}
{\bf u}={\rm e}^{-{\rm j}(k_x x+k_y y)} {\bf Y}_2 {\bf E} {\bf C}
\end{eqnarray}
Here, the matrix ${\bf E}$ and the vector ${\bf C}$ are provided in Eq.(\ref{eq:matrix_e})-(\ref{eq:vec_c}). The non-zero elements of the coefficient matrix ${\bf Y}_2$ are provided in \ref{app:case2_exp}.

\subsection{Incident case 3}
When $k_y=k_x= 0$, the solid phase strain ${\bf \Omega}_s={\bf 0}$. Subsequently, the displacement ${\bf u}=[u_x^s,u_y^s,u_z^s,u_x^f,u_y^f,u_z^f]^T$ is
\begin{eqnarray}
{\bf u}={\rm e}^{-{\rm j}(k_x x+k_y y)} {\bf Y}_3 \hat{{\bf E}} \hat{{\bf C}}
\end{eqnarray}
where
\begin{eqnarray}
&\hat{{\bf E}} =diag({\rm e}^{-{\rm j} k_{1z}^{m} z},{\rm e}^{{\rm j} k_{1z}^{m} z},{\rm e}^{-{\rm j} k_{2z}^{m} z},{\rm e}^{{\rm j} k_{2z}^{m} z},0,0)\\
&\hat{{\bf C}} =[C_1,C_2,C_3,C_4,0,0]^T
\end{eqnarray}
Here, the matrix $\hat{{\bf E}}$ is a $6\times6$ diagonal matrix. The non-zero elements of the coefficient matrix ${\bf Y}_3$ are given in \ref{app:case3_exp}.

Once the poroelastic displacements in the porous media are obtained, their force components can be solved using the stress-strain relations in Biot theory \cite{Allard2009}. The poroelastic field is revealed subsequently.

\section{Discussions}

\subsection{Comparison with the 3D results by Zhou}

\begin{figure}[!htbp]
     \centering
     \includegraphics[width=0.8\textwidth]{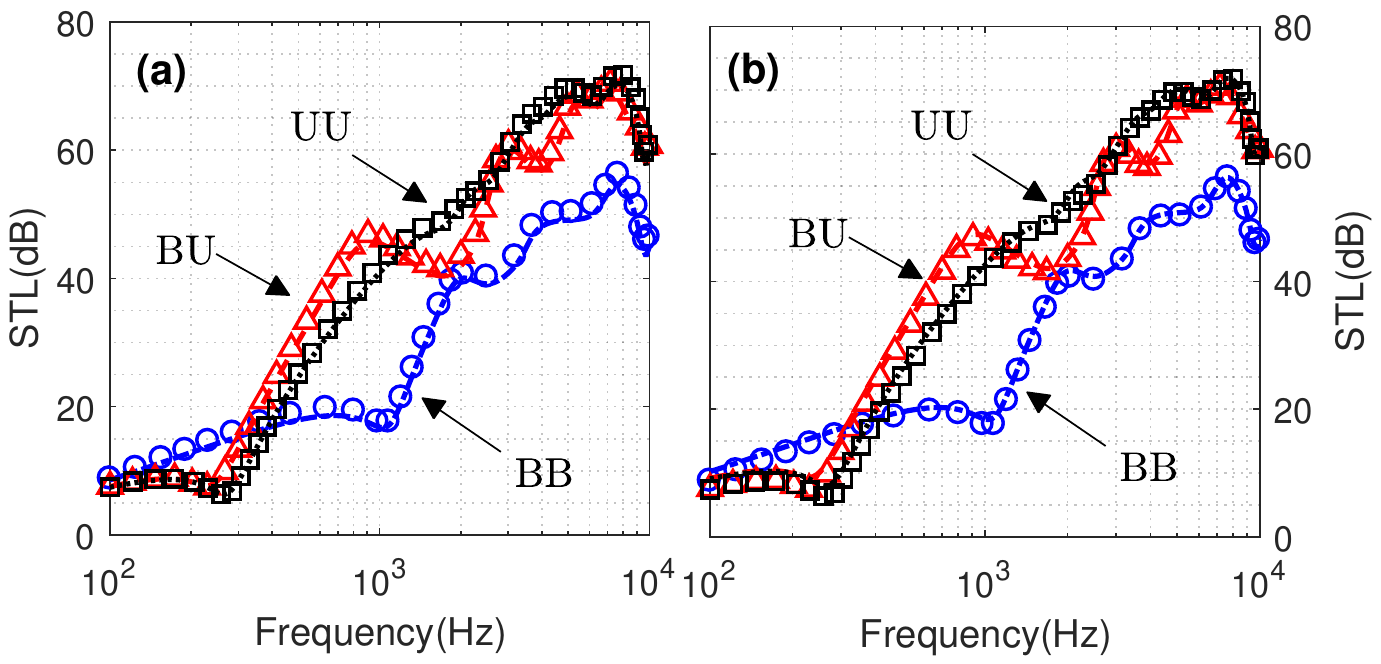}
     \caption{Comparison of the random STL results with Zhou: (a) the results here (the lines) versus Zhou's results (the marks); (b) the results here exclude division-by-zero cases (the lines) versus Zhou's results (the marks)}\label{fig:cmp_with_zhou}
 \end{figure}

The problems proposed by Zhou \cite{Zhou2013JoSaV} (a 3D double panel with porous lining under different boundary conditions) are solved using the field expressions obtained here. Their boundary formulations and parameters can be found in \cite{Zhou2013JoSaV}. The results obtained here versus Zhou's results, in the Mach number $M=0$ case, are provided in Fig.\ref{fig:cmp_with_zhou}(a). It is found that Zhou overestimated the results slightly, though the same trends are shared, owing to the incomplete 3D poroelastic field expressions in \cite{Zhou2013JoSaV}. His field expressions, when $k_x=k_1 {\rm cos}\varphi_1 {\rm cos}\theta_1=0$ (the same definitions and notations as \cite{Zhou2013JoSaV}), are invalid and lead to division by zero. As $k_1\ne0$, $\varphi_1\in[\pi/10,\pi/2]$ and $\theta_1\in[0,2\pi]$, these division-by-zero cases lead to $\varphi_1=\pi/2$, or $\theta_1=\pi/2,3\pi/2$. These three incident cases, which are not properly discussed by Zhou, should be correctly solved to obtain the STL.

To study the influence of the division-by-zero cases, the results here with the division-by-zero cases excluded, versus Zhou's results, are provided in Fig.\ref{fig:cmp_with_zhou}(b). It shows the consistency of the two results is more satisfactory than in Fig.\ref{fig:cmp_with_zhou}(a). These results confirm that, the results by Zhou, though amply acceptable, fail to reveal the division-by-zero cases and lead to an overestimate of the STL.
 
 \begin{figure}[!htbp]
     \centering
     \includegraphics[width=.8\textwidth]{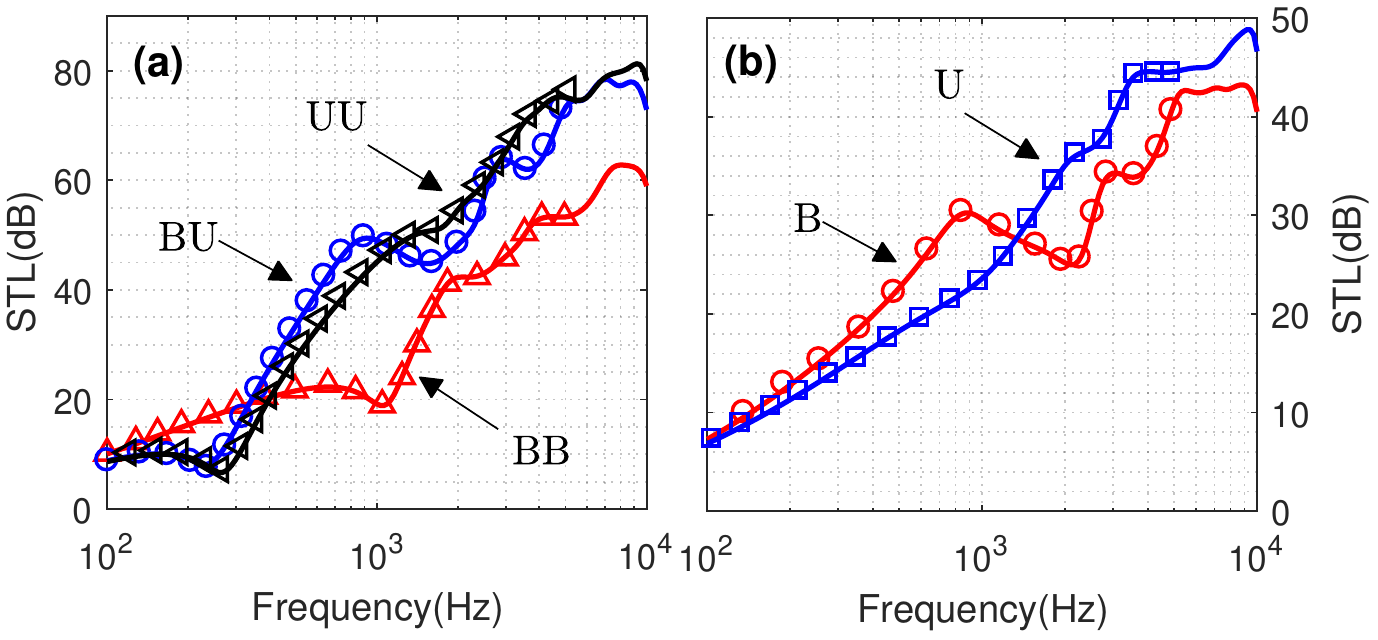}
     \caption{Comparison of the random STL results with Bolton: the results here (the lines) versus Bolton's results (the marks) (a) the BB, BU, and UU cases (b) the B and U cases}\label{fig:val_with_bolton}
 \end{figure}

\subsection{Validation with the 2D results by Bolton}

In his paper \cite{Bolton1996JoSaV}, Bolton obtained 2D poroelastic field expressions and studied the random STL of a 2D double panel with porous lining under different boundary conditions. These 2D results are compared with the results obtained using the field expressions provided here (with $k_y=0$) in Fig.\ref{fig:val_with_bolton} (the same configurations and parameters in \cite{Bolton1996JoSaV} are used). All the five boundary cases are in good consistency. These results show the correctness of the 3D field expressions obtained here.

\subsection{Verification of the division-by-zero cases}
The division-by-zero cases, owing to incomplete expressions \cite{Zhou2013JoSaV}, are tested to verify the field expressions got here. Three 3D oblique incident conditions are identified here, which are $\varphi_1=\pi/2$, $\theta_1=\pi/2$ or $\theta_1=3\pi/2$ (where the elevation angle $\varphi_1$ and the azimuth angle $\theta_1$ are defined in \cite{Zhou2013JoSaV}). However, as $\theta_1=\pi/2$ or $3\pi/2$ correspond to the 2D plane incident cases which were validated using Bolton’s results, only $\varphi_1=\pi/2$, i.e., the normal incident case, is verified here. 

\begin{figure}
     \centering
     \includegraphics[width=0.6\textwidth]{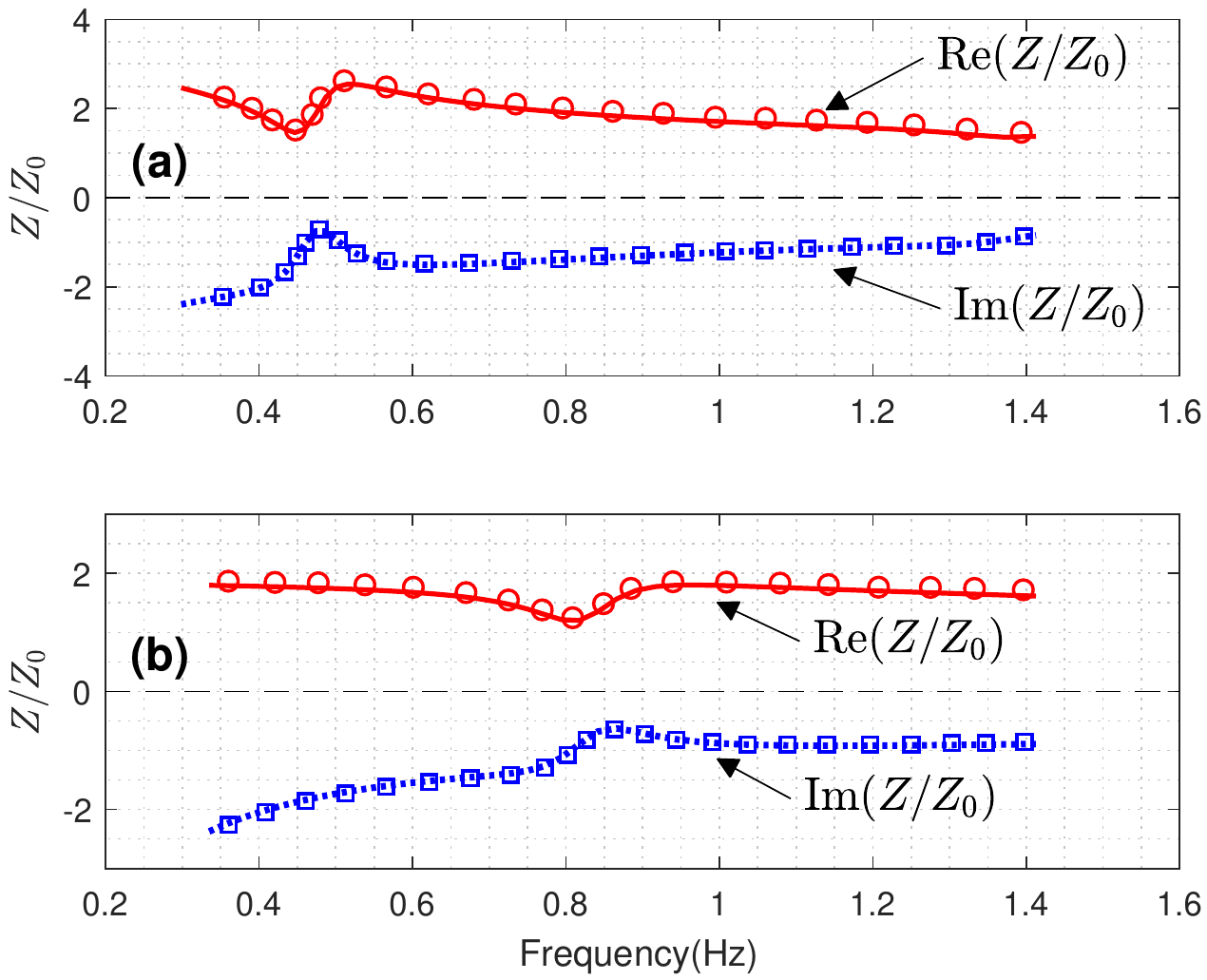}
     \caption{Comparison of the normalized surface impedance at normal incidence when the porous layer thickness is (a) 10 cm (b) 5.6 cm (the lines - the results obtained here; the marks - Allard's results)}\label{fig:valNmCWithAllard}
\end{figure}

A problem proposed by Allard \cite{Allard2009}, which is a poroelastic layer bonded on to a rigid impermeable wall in a normal acoustic field, is discussed to test the normal incidence solution (the same configuration and parameters in \cite{Allard2009} are used). Result comparisons with Allard are provided in Fig.\ref{fig:valNmCWithAllard}. As shown, the consistency between the results got here and Allard’s results is satisfactory. This consistency confirms the effectiveness of the 3D poroelastic field expressions got here.

\section{Conclusions}\label{stn:conclusion}
The correct 3D poroelastic field expressions, under monochromatic harmonic incidence, were revealed here. Result comparisons of these field expressions with former expressions showed that the results got formerly, though acceptable, overestimated the random STL slightly and led to division-by-zero. The division-by-zero cases and validations with 2D theoretical results were subsequently tested. Their result comparisons were all satisfactory and confirmed the soundness of the 3D field expressions got here. As a practicable and reliable tool, these 3D field expressions can benefit to understand the acoustic properties of poroelastic materials.

\section*{Acknowledgements}
This work is supported by the National Natural Science Foundation of China (NSFC) No.11572137. The authors wish to thank Professor J. Stuart Bolton, Professor Yu Liu and Dr. Jie Zhou for their warm-hearted discussions.

\appendix

\section{The coefficient matrix of incident case 1}\label{app:case1_exp}
The non-zero elements of matrix ${\bf Y}_1$ are
\begin{eqnarray}
&Y_1(1,1)=Y_1(1,2)=\frac{{\rm j} k_x}{k_1^2},\ Y_1(1,3)=Y_1(1,4)=\frac{{\rm j} k_x}{k_2^2}\notag
\\
& Y_1(1,5)=-\frac{{\rm j} k_x k_{3z}}{k_3^2 k_y},\ Y_1(1,6)=- Y_1(1,5)\notag
\\
&Y_1(2,1)=Y_1(2,2)=\frac{{\rm j} k_y}{k_1^2},\ Y_1(2,3)=Y_1(2,4)=\frac{{\rm j} k_y}{k_2^2}\notag
\\
&Y_1(2,5)=-\frac{{\rm j} k_{3z}}{k_3^2},\ Y_1(2,6)=-Y_1(2,5)\notag
\\
&Y_1(3,1)={\rm j}\frac{k_{1z}}{k_1^2},\ Y_1(3,2)=-Y_1(3,1)\notag \\
&Y_1(3,3)={\rm j}\frac{k_{2z}}{k_2^2},\ Y_1(3,4)=-Y_1(3,3)\notag\\
&Y_1(3,5)=\frac{{\rm j}(k_x^2+k_y^2)}{k_3^2 k_y},\ Y_1(3,6)=Y_1(3,5)\notag\\
&Y_1(4,1)=Y_1(4,2)=\frac{{\rm j} k_x}{k_1^2}b_1,\ Y_1(4,3)=Y_1(4,4)=\frac{{\rm j} k_x}{k_2^2}b_2\notag
\\
&Y_1(4,5)=-\frac{{\rm j} k_x k_{3z}}{k_3^2 k_y}g,\ Y_1(4,6)=- Y_1(4,5)\notag
\\
&Y_1(5,1)=Y_1(5,2)=\frac{{\rm j} k_y}{k_1^2}b_1,\ Y_1(5,3)=Y_1(5,4)=\frac{{\rm j} k_y}{k_2^2}b_2\notag
\\
&Y_1(5,5)=-\frac{{\rm j} k_{3z}}{k_3^2}g,\ Y_1(5,6)=-Y_1(5,5)\notag
\\
&Y_1(6,1)={\rm j}\frac{k_{1z}}{k_1^2}b_1,\ Y_1(6,2)=-Y_1(6,1)\notag \\
&Y_1(6,3)={\rm j}\frac{k_{2z}}{k_2^2}b_2,\ Y_1(6,4)=-Y_1(6,3)\notag\\
&Y_1(6,5)=\frac{{\rm j}(k_x^2+k_y^2)}{k_3^2 k_y}g,\ Y_1(6,6)=Y_1(6,5)
\end{eqnarray}

\section{The coefficient matrix of incident case 2}\label{app:case2_exp}
The non-zero elements of matrix ${\bf Y}_2$ are
\begin{eqnarray}
&Y_2(1,1)=Y_2(1,2)=\frac{{\rm j} k_x}{k_1^2},\ Y_2(1,3)=Y_2(1,4)=\frac{{\rm j} k_x}{k_2^2}\notag
\\
& Y_2(1,5)=\frac{{\rm j} k_{3z}}{k_3^2},\ Y_2(1,6)=-Y_2(1,5)\notag
\\
&Y_2(2,1)=Y_2(2,2)=\frac{{\rm j} k_y}{k_1^2},\ Y_2(2,3)=Y_2(2,4)=\frac{{\rm j} k_y}{k_2^2}\notag
\\
&Y_2(2,5)=\frac{{\rm j} k_y k_{3z}}{k_3^2 k_x},\ Y_2(2,6)=-Y_2(2,5)\notag
\\
&Y_2(3,1)={\rm j}\frac{k_{1z}}{k_1^2},\ Y_2(3,2)=-Y_2(3,1)\notag \\
&Y_2(3,3)={\rm j}\frac{k_{2z}}{k_2^2},\ Y_2(3,4)=-Y_2(3,3)\notag\\
&Y_2(3,5)=-\frac{{\rm j}(k_x^2+k_y^2)}{k_3^2 k_x},\ Y_2(3,6)=Y_2(3,5)\notag\\
&Y_2(4,1)=Y_2(4,2)=\frac{{\rm j} k_x}{k_1^2}b_1,\ Y_2(4,3)=Y_2(4,4)=\frac{{\rm j} k_x}{k_2^2}b_2\notag
\\
&Y_2(4,5)=\frac{{\rm j} k_{3z}}{k_3^2}g,\ Y_2(4,6)=- Y_2(4,5)\notag
\\
&Y_2(5,1)=Y_2(5,2)=\frac{{\rm j} k_y}{k_1^2}b_1,\ Y_2(5,3)=Y_2(5,4)=\frac{{\rm j} k_y}{k_2^2}b_2\notag
\\
&Y_2(5,5)=\frac{{\rm j} k_y k_{3z}}{k_3^2 k_x}g,\ Y_2(5,6)=-Y_2(5,5)\notag
\\
&Y_2(6,1)={\rm j}\frac{k_{1z}}{k_1^2}b_1,\ Y_2(6,2)=-Y_2(6,1)\notag \\
&Y_2(6,3)={\rm j}\frac{k_{2z}}{k_2^2}b_2,\ Y_2(6,4)=-Y_2(6,3)\notag\\
&Y_2(6,5)=-\frac{{\rm j}(k_x^2+k_y^2)}{k_3^2 k_x}g,\ Y_2(6,6)=Y_2(6,5)
\end{eqnarray}

\section{The coefficient matrix of incident case 3}\label{app:case3_exp}
The non-zero elements of matrix ${\bf Y}_3$ are
\begin{eqnarray}
&Y_3(3,1)=\frac{{\rm j}}{k_1},\ Y_3(3,2)=-Y_3(3,1)\notag
\\
&Y_3(3,3)=\frac{{\rm j}}{k_2},\ Y_3(3,4)=-Y_3(3,3)\notag
\\
&Y_3(6,1)=\frac{{\rm j}}{k_1}b_1,\ Y_3(6,2)=-Y_3(6,1)\notag
\\
&Y_3(6,3)=\frac{{\rm j}}{k_2}b_2,\ Y_3(6,4)=-Y_3(6,3)
\end{eqnarray}
 

\bibliography{mybibfile}

\end{document}